\documentclass[conference]{IEEEtran}
\IEEEoverridecommandlockouts

\usepackage{cite}
\usepackage{amsmath,amssymb,amsfonts}
\usepackage{algorithmic}
\usepackage{graphicx}
\usepackage{textcomp}
\usepackage{xcolor}
\usepackage{listings}
\usepackage{xcolor}
\usepackage[hidelinks]{hyperref}
\usepackage{subcaption}

\definecolor{darkgreen}{rgb}{0,0.7,0}

\lstdefinestyle{pythonstyle}{
  language=Python,
  basicstyle=\ttfamily\footnotesize,
  keywordstyle=\color{red},
  commentstyle=\color{gray},
  stringstyle=\color{orange},
  showstringspaces=false,
  breaklines=true,
  xleftmargin=5pt,
  xrightmargin=5pt,
  numbers=none,
  numberstyle=\tiny\color{gray},
  tabsize=4,
  morekeywords={create_cwl_app},
    frame=tblr,  
    rulecolor=\color{black},
    captionpos=b,  
    breaklines=true,
    breakatwhitespace=false,
    tabsize=2,
    abovecaptionskip=1pt,
    belowcaptionskip=1pt
}

\lstdefinestyle{yamlstyle}{
    language=, 
    basicstyle=\ttfamily\footnotesize,
    keywordstyle=\color{blue},
    commentstyle=\color{gray},
    stringstyle=\color{red},
    showstringspaces=false,
  breaklines=true,
  xleftmargin=5pt,
  xrightmargin=5pt,
  numbers=none,
    numberstyle=\tiny\color{gray},
    tabsize=4,
    morekeywords={cwlVersion, class, baseCommand, inputs, type, default, inputBinding, position, stdout, outputs, output, doc, size, sepia, radius, final_output, requirements, input_image, steps, run, in, out, valueFrom, outputSource,arguments, expressionLib},
    frame=tblr,  
    rulecolor=\color{black},
    captionpos=b,  
    breaklines=true,
    breakatwhitespace=false,
    tabsize=2,
    abovecaptionskip=1pt,
    belowcaptionskip=1pt
}

\lstdefinestyle{bashstyle}{
    language=,
    basicstyle=\ttfamily\footnotesize,
    keywordstyle=\color{green},
    numbers=none,
    rulecolor=\color{black},
    captionpos=b,
    breakatwhitespace=false,
    tabsize=2,
    abovecaptionskip=1pt,
    belowcaptionskip=1pt,
}

\def\BibTeX{{\rm B\kern-.05em{\sc i\kern-.025em b}\kern-.08em
    T\kern-.1667em\lower.7ex\hbox{E}\kern-.125emX}}

\usepackage{caption}
\begin{document}

\title{Parsl+CWL: Towards Combining the Python and CWL Ecosystems}


\makeatletter
\newcommand{\linebreakand}{%
  \end{@IEEEauthorhalign}
  \hfill\mbox{}\par
  \mbox{}\hfill\begin{@IEEEauthorhalign}
}
\makeatother

\author{\IEEEauthorblockN{
Nishchay Karle\IEEEauthorrefmark{1}, 
Ben Clifford\IEEEauthorrefmark{1}, 
Yadu Babuji\IEEEauthorrefmark{1}\IEEEauthorrefmark{2}, 
Ryan Chard\IEEEauthorrefmark{2},
Daniel S. Katz\IEEEauthorrefmark{2} and
Kyle Chard\IEEEauthorrefmark{1}\IEEEauthorrefmark{2}}
\IEEEauthorblockA{\IEEEauthorrefmark{1}
Department of Computer Science,  University of Chicago, 
Chicago, USA\\
\IEEEauthorrefmark{2}
Argonne National Laboratory, Lemont, USA\\
\IEEEauthorblockA{\IEEEauthorrefmark{3}NCSA \& CS \& iSchool, University of Illinois Urbana-Champaign, Urbana, IL, USA }}

}

\maketitle
\begin{abstract}
The Common Workflow Language (CWL) is a widely adopted language for defining and sharing computational workflows. It is designed to be independent of the execution engine on which workflows are executed. In this paper, we describe our experiences integrating CWL with Parsl, a Python-based parallel programming library designed to manage execution of workflows across diverse computing environments. 
We propose a new method that converts CWL CommandLineTool definitions into Parsl apps, enabling Parsl scripts to easily import and use tools represented in CWL. We describe a Parsl runner that is capable of executing a CWL CommandLineTool directly. We also describe a proof-of-concept extension to support inline Python in a CWL workflow definition, enabling seamless use in Parsl's Python ecosystem. We demonstrate the benefits of this integration by presenting example CWL CommandLineTool definitions that show how they can be used in Parsl, and comparing performance of executing an image processing workflow using the Parsl integration and other CWL runners.
\end{abstract}

\section{Introduction}


Scientific workflows are essential for automating complex computational tasks, enabling reproducibility, portability, and scalability of science applications. The pervasiveness of scientific workflows has led to the development of hundreds of workflow management systems that support the development and execution of workflows; however, most workflow systems are not interoperable and thus workflows developed using one workflow system are not usable in another. The Common Workflow Language (CWL) attempts to overcome this obstacle via a common workflow description that can be interpreted by many workflow systems. CWL has been widely adopted, and users have created a rich ecosystem of CWL workflows and tools. Importantly, the community has undertaken the significant effort to describe tools (i.e., applications and scripts) in CWL, including specifying input/output formats, command line invocation arguments, and environment requirement. These tool definitions are a critical step, irrespective of workflow system, in being able to programatically execute a tool as part of a workflow. 

CWL is designed such that workflow definitions are independent of the \textit{CWL runner}---the workflow system that executes the workflow. Many CWL runners have been implemented, such as CWLTool~\cite{CWLtool}, Toil~\cite{Vivian2017}, Arvados~\cite{Arvados}, and previously Cromwell~\cite{Cromwell}.  These runners provide distinct capabilities and thus have distinct user communities. An advantage of CWL is that workflows can be easily ported between these different runners enabling users to choose a runner that best matches their requirements. 

Here, we describe work towards integrating CWL and Parsl~\cite{babuji19parsl}---a parallel programming library designed to enable parallel Python execution across different computing resources, from local clusters to cloud platforms. Parsl's dataflow model allows for intuitive definition of workflows directly in the Python programming language. Parsl's flexible execution framework enables scalable and efficient execution of workflows across many computing platforms, particularly at scale on large HPC systems. These capabilities make Parsl a potentially valuable runner for CWL workflows. 


Our integration of CWL and Parsl aims to make it possible for Parsl developers to programmatically import and invoke CWL-defined tools directly in Python programs. Importantly, this integration removes the need for Parsl developers to manually specify and maintain tool definitions in Parsl's Python-based representation. 
Importing CWL tools directly into Parsl programs enables a richer programmatic approach to composing workflows that might include CWL tools,
existing tools represented in Parsl, pure Python functions, and program logic written in Python to manage the execution of the workflow. We further  implement a proof-of-concept CWL CommandLineTool runner that enables Parsl to execute CommandLineTools using Parsl's robust, scalable, and performant executors. 



We see several benefits of integrating CWL and Parsl:
\begin{itemize}
  \item \textbf{Portability}: CWL provides a common way to describe tools, ensuring that they can be executed on different platforms.
  \item \textbf{CWL ecosystem}: There are many CWL CommandLineTool definitions that describe input and output formats, command line interfaces, and environment requirements that can be used directly in Parsl without requiring developers to recreate these definitions in Python. 
  \item \textbf{Scalability and Performance}: Parsl's runtime engine and various executors efficiently manage resources, allowing workflows to scale from personal computers to high-performance computing clusters.
  \item \textbf{Familiarity/Productivity}: Python is arguably the lingua franca of Science.
  Our CWL and Parsl integration enables the composition of workflows in Python while leveraging the curated
  tool definitions from the CWL ecosystem.
\end{itemize}


Our experiences integrating Parsl and CWL highlighted the challenges of supporting CWL expressions---snippets of code written in JavaScript embedded in CWL YAML workflow definitions---given Parsl's Python-based environment. To overcome this mismatch we propose a prototype extension to CWL to support inline Python expressions in CWL workflow definitions.
Inline Python allows for dynamic logic within workflows to be described entirely in Python. 


This paper is structured as follows. \S\ref{sec:background} describes CWL and Parsl. \S\ref{sec:cwlparsl} outlines our extensions to Parsl to import and run CWL CommandLineTools. 
\S\ref{sec:example} presents an example workflow and shows how the CWL workflow can be implemented in Parsl using our integration. 
\S\ref{sec:inlinepython} describes how we support Python expressions in CWL. \S\ref{sec:evaluation} compares the performance of our integration and Python expressions with other CWL runners. Finally, \S\ref{sec:relatedwork} describes related work and \S\ref{sec:summary} summarizes our contributions.  

The code described in this paper is openly available on GitHub (\url{https://github.com/Parsl/cwl-parsl}) under the Apache-2 license.

\section{Background}
\label{sec:background}

Here we describe the foci of our integration: CWL and Parsl.

\subsection{Common Workflow Language (CWL)}

The Common Workflow Language (CWL)~\cite{Crusoe2022} is an open specification that is designed to address the challenges of reproducibility and interoperability in scientific research. CWL achieves this goal by providing a common specification for workflows to ensure they can be shared and reused irrespective of the underlying workflow engine used. 

%

CWL has two main abstractions: CommandLineTools and Workflows. \textit{CommandLineTool} definitions, written in YAML, outline the interface to a command line tool (e.g., an application, script, or anything that can be invoked via the command line). The tool definition describes the input arguments, environment, and output files. The definition can then be used to invoke the command line tool, given suitable input arguments. The tool definition format allows definitions to be shared across workflows and referenced from registries. The community has invested significant effort cataloging tools and sharing definitions~\cite{Goble2021}. 

CommandLineTool definitions are used in a CWL \textit{Workflow} definition, also written in YAML. The workflow links together CommandLineTools by specifying the exchange of input/output between tools. Importantly, while the workflow describes the various steps (and their input/outputs), execution of the tools is determined by dependencies rather than the order they are specified in the workflow definition. 
CWL supports software container technologies (e.g., Docker) to abstract execution environments.  

While the CWL specification is a static representation of a workflow in YAML, there are many situations in which dynamic decisions need to be made as a workflow progresses, for example, to select a specific CommandLineTool to execute based on the output of a previous CommandLineTool or to modify arguments passed between CommandLineTools. CWL provides several built-in methods for common manipulations and also supports arbitrary \textit{expressions}---snippets of JavaScript code that are evaluated during workflow execution. 

A CWL workflow is executed by a CWL \textit{runner}. The runner is responsible for managing the invocation of the CommandLineTools, determining when they can be executed, composing the execution command, monitoring execution and determining success or failure, and managing the exchange of data between CommandLineTools. \textit{cwltool}~\cite{CWLtool} is the reference implementation of CWL runner. It is implemented in Python and maintained by the CWL community. cwltool is able to validate CWL descriptions, parsing the representation and ensuring that it is compliant with the CWL specification. It can also execute the workflow with user-supplied input arguments and a workflow definition. \textit{toil-cwl-runner} is a Python-based CWL runner built on the Toil workflow engine. It validates CWL descriptions for compliance with the CWL specification and can execute workflows on both single-node setups and distributed cloud environments, using user-supplied inputs and workflow definitions.

\subsection{Parsl}

Parsl~\cite{babuji19parsl} is a parallel programming library for Python. Parsl allows developers to write programs entirely in Python and Parsl then manages execution of those programs across diverse computing resources. Parsl abstracts the complexities inherent in parallel computing by providing
a straightforward functional programming model at the task
level while maintaining procedural Python code for the wider program and task and data dependencies.

In Parsl, developers annotate Python functions as \texttt{apps} to specify that 
they can be executed concurrently. When an \texttt{app} is invoked, a \textit{Future} is 
returned that tracks the asynchronous execution of the \textit{app}.
Dataflow is implicitly specified when a Future from one app is passed as input to another app.
Parsl dynamically generates a task dependency graph and then maps the graph to available resources for execution, exploiting parallelism where possible by managing the
creation of data objects and ensuring that app dependencies are met.

Parsl implements an extensible plugin model for its runtime execution system called \textit{Executors}. 
Executors implement  Python's \texttt{concurrent.futures.Executor} class and are responsible for executing a task and returning a future to the calling program.
Parsl supports standard implementations of the \texttt{concurrent.futures.Executor} class, such as the \textit{ThreadPoolExecutor}. It also includes several Parsl-specific Executors, such as  the \textit{HighThroughputExecutor}, and interfaces with other community Executors, such as \textit{TaskVineExecutor} and \textit{RadicalPilotExecutor}. 

The most commonly used Executor,  the \textit{HighThroughputExecutor} (HTEX), employs a pilot job model to manage the execution of tasks on a paralel or distributed computer.  This model introduces an abstraction layer that decouples task submission from resource allocation, thereby enabling efficient utilization of available computational resources. In the pilot job model, a placeholder job---referred to as a pilot---is submitted to a batch scheduler. Tasks are then executed on the pilot job without interfacing with the batch scheduler.  The pilot job model is particularly advantageous in environments with high variability in job queue time. 

Parsl also implements an extensible \textit{Provider} interface that facilitates the negotiation of computing resources from a range of batch systems, public clouds, and container orchestration systems like Kubernetes. \textit{Providers} in Parsl are responsible for managing the lifecycle of compute resources, including provisioning, monitoring, and deprovisioning resources, thus enabling automatic scaling to match the needs of the workflow at runtime. This abstraction of the resource management system, allows Parsl to run on local clusters, cloud environments, and supercomputers with minimal configuration changes.

Parsl interacts with many other tools, such as TaskVine as mentioned before, as well as Globus~\cite{chard14globus}. Parsl is used in a range of scientific applications, has been shown to scale to some of the largest supercomputers, and is used as the basis for building other services, such as Globus Compute~\cite{chard20funcx}. 

\section{Integrating CWL and Parsl}
\label{sec:cwlparsl}

Our integration of CWL and Parsl focuses on two primary areas: 1) importing CWL CommandLineTool definitions in Parsl, and 2) implementing a prototype CWL Parsl runner to execute CommandLineTools via Parsl.


\subsection{Importing Tool Definitions}

Given a CWL CommandLineTool definition, we seek to integrate the tool into a Parsl program such that it can be executed like any other Parsl app and therefore interwoven seamlessly in the program.  The challenge here is that CWL CommandLineTool definitions are written in a YAML format while Parsl apps are represented as Python functions. 

To overcome this difference we introduce a new Parsl app: \textit{CWLApp}. The CWLApp reads a CWL CommandLineTool definition and transparently creates a Parsl \textit{BashApp} that is configured to run the CWL CommandLineTool. 
The process of creating a CWLApp requires only that the developer specifies a file containing the CWL CommandLineApp.  The CWLApp will read the definition and populate the input/output definition for the Parsl app. 
The CWLApp is callable as a Python function, allowing users to execute the tool by passing the required input arguments. These inputs, combined with the definitions from the CWL CommandLineTool, are used to construct and then execute the command. 



The inputs specified in the CWL CommandLineTool are represented as keyword arguments in the CWLApp. When invoked, these arguments are processed according to the specifications in the CWL CommandLineTool definition. Prefixes and positions defined in the CommandLineTool's \textit{inputBinding} definition are matched with arguments
at runtime. Any inputs that are of type ``File'' are converted into Parsl's \textit{File} type, which facilitate access regardless of the location where the app is executed.

The outputs specified in the CommandLineTool are managed as described in the definition. Both \textit{stdout} and \textit{stderr} are directed to their designated files. For any new file created during execution, a Parsl \textit{DataFuture} object is created and returned. These DataFuture objects can then be passed as file inputs to other Parsl apps (including CWLApps) without needing to wait for the files to be available.

CWLApps can be created once and imported, allowing them to be reused across workflows. They can also be created in a Python module and then imported directly into other Python programs. 

Listing~\ref{lst:echocwl} shows a CommandLineTool definition for the Linux echo command. It describes the required input argument, a string called message, and the output stdout file that is to be produced. Listing~\ref{lst:echo-parsl} shows how the CommandLineTool definition is used to create a CWLApp in Parsl and how the CommandLineTool can then be executed with Parsl.


\lstset{style=yamlstyle}
\begin{lstlisting}[caption={CWL CommandLineTool definition for ``echo''}, label=lst:echocwl]
cwlVersion: v1.2

class: CommandLineTool

baseCommand: echo

inputs:
  message:
    type: string
    default: "Hello World"
    inputBinding:
      position: 1

outputs:
  output:
    type: stdout

stdout: hello.txt

\end{lstlisting}

\lstset{style=pythonstyle}
\begin{lstlisting}[caption={An example Parsl program that first loads a Parsl configuration, loads the CWL CommandLineTool definition from the echo.cwl file, executes the CommandLineTool using Parsl, waits for the task to complete, and prints the contents of the output file.}, label=lst:echo-parsl]
from parsl.configs.local_threads import config
from parsl_cwl.cwl_app import CWLApp

parsl.load(config)

echo = CWLApp("echo.cwl")

future = echo(
    message="Hello, World!",
    stdout="hello.txt",
)

# Wait for the future before reading the output
future.result()

with open("hello.txt", "r") as f:
    print(f.read())
\end{lstlisting}

\subsection{Parsl CWL CommandLineTool Runner}

We now describe how Parsl can act as a CWL runner to execute a single CWL CommandLineTool definition. This integration enables users to leverage Parsl's scalability and performance when running a CommandLineTool on high performance parallel and distributed systems. 
Using the Parsl CWL CommandLineTool runner, a user can simply pass a CWL CommandLineTool definition to run with the Parsl CWL runner command: \textit{parsl-cwl}. Currently, parsl-cwl can only execute CWL CommandLineTools directly; in the future we will extend this integration to support Workflow definitions. 

Here, we show an example of how a CWL CommandLineTool can be executed by specifying the CommandLineTool definition (\texttt{echo.cwl}), defining the Parsl configuration to use (\texttt{config.yml}), and specifying inputs either as command line arguments or as a YAML file (\texttt{inputs.yml}):

\lstset{style=bashstyle}
\begin{lstlisting}
$ parsl-cwl config.yml echo.cwl inputs.yml
\end{lstlisting}
\begin{lstlisting}
$ parsl-cwl config.yml echo.cwl --message='Hello'
\end{lstlisting}

We adopt a YAML-based Parsl configuration to match the CWL ecosystem. Specifically, we use the format defined in the TaPS benchmark suite~\cite{TaPS} to specify key configuration options. The configuration includes the executor and provider, as well as various options such as number of nodes, number of workers per node, accelerators to be used, and Python or container environment.


\section{Example Workflow}
\label{sec:example}

We now present an example image processing workflow written in CWL and ported to Parsl using our integration.

\subsection{Image Processing Workflow in CWL}

Listing~\ref{lst:CWL_Image_Processing_Example} shows the CWL Workflow definition.
%
The workflow encodes a sequence of image processing tasks: resizing an image, applying a sepia filter, and blurring the image. Each stage is implemented as a separate CWL \textit{CommandLineTool}. The workflow takes an image as input, and produces a blurred and possibly sepia-filtered image as output. 
The three stages of the workflow are: 

\begin{enumerate}
    \item \textbf{Image Resizing (\texttt{resize\_image.cwl})}:
This stage takes an input image and resizes it to the specified dimensions. The target size is provided as an input parameter, allowing for flexible resizing according to user requirements.

\item 
\textbf{Image Filtering (\texttt{filter\_image.cwl})}:
This stage applies a sepia filter to the resized image. The filter is controlled by a boolean parameter, enabling a user to apply the sepia effect as needed.

\item
\textbf{Image Blurring (\texttt{blur\_image.cwl})}:
The final stage blurs the filtered image using a specified blur radius. This step allows a user to soften the image, with the degree of blurring controlled by an input parameter. The final blurred image is saved as \texttt{blurred.png}.
\end{enumerate}

The workflow takes four input arguments.
\begin{itemize}
    \item \texttt{input\_image}: The original image file to be processed.
    \item \texttt{size}: The dimensions to which the image should be resized.
    \item \texttt{sepia}: A boolean flag indicating whether the sepia filter should be applied.
    \item \texttt{radius}: The radius used for the blur operation.
\end{itemize}

The workflow produces one output file: 
\begin{itemize}
    \item \texttt{final\_output}: The final blurred image, output from the \texttt{blur\_image} stage.
\end{itemize}

\lstset{style=yamlstyle}
\begin{lstlisting}[caption={CWL Image Processing Workflow}, label={lst:CWL_Image_Processing_Example}]
cwlVersion: v1.2

class: Workflow

doc: This CWL workflow processes images by performing a series of tasks - resizing, filtering, and blurring

requirements:
  - class: StepInputExpressionRequirement

inputs:
  input_image:
    type: File
    doc: The original image to be processed

  size:
    type: int
    doc: The target sizeXsize for resizing

  sepia:
    type: boolean
    doc: Whether to apply the filter

  radius:
    type: int
    doc: The amount of blur to apply

outputs:
  final_output:
    type: File
    outputSource: blur_image/output_image

steps:
  resize_image:
    run: resize_image.cwl
    in:
      input_image: input_image
      size: size
      output_image:
        valueFrom: "resized.png"
    out: [output_image]

  filter_image:
    run: filter_image.cwl
    in:
      input_image: resize_image/output_image
      sepia: sepia
      output_image:
        valueFrom: "filtered.png"
    out: [output_image]

  blur_image:
    run: blur_image.cwl
    in:
      input_image: filter_image/output_image
      radius: radius
      output_image:
        valueFrom: "blurred.png"
    out: [output_image]
\end{lstlisting}

\subsection{Image Processing Workflow in Parsl}

Listing~\ref{lst:parslworkflow} shows the image processing workflow from Listing~\ref{lst:CWL_Image_Processing_Example} implemented in Parsl by importing CWL CommandLineTool definitions and orchestrating the execution of each CommandLineTool in sequence. We present this example to illustrate how the CWL CommandLineTools can be integrated in a Parsl program and to highlight the advantages of using a Python interface to specify the control and data flow between steps. 
In this example, we take a modular and Pythonic approach by defining a function representing the three sequential steps, which can then be used repeatedly (e.g., in a loop or in another function) to process multiple images concurrently.  Parsl will derive a DAG of the individual tasks and execute them in an interleaved fashion when their dependencies are met (i.e., it will not wait for the three steps to be completed before running a step from another loop).

\lstset{style=pythonstyle}
\begin{lstlisting}[caption={Python Script Example}, label={lst:parslworkflow}]
import glob

from myconfigs import perlmutter_config
from parsl_cwl.cwl_app import CWLApp


parsl.load(perlmutter_config)

resize_image = CWLApp("resize_image.cwl")
filter_image = CWLApp("filter_image.cwl")
blur_image = CWLApp("blur_image.cwl")

def process_img(image: str) -> Future:
    resized_img_future = resize_image(
        input_image=image,
        size=1024,
    )

    filtered_img_future = filter_image(
        input_image=resized_img_future.outputs[0],
        sepia=True,
    )

    blurred_img_future = blur_image(
        input_image=filtered_img_future.outputs[0],
        radius=1,
    )

    return blurred_img_future


final_imgs = [process_img(img) for img in glob.glob(r'**/*.png', recursive=True)]

concurrent.futures.wait(
    final_imgs, return_when=concurrent.futures.ALL_COMPLETED
)
\end{lstlisting}

We describe in listing \ref{lst:parslworkflow} the various parts of the Parsl program to illustrate how one can use the CWL integration.

\textbf{Configuration and Executor Setup}:
We first configure a \texttt{HighThroughputExecutor} to manage execution of the workflow. In this case, we load a configuration for the Perlmutter Supercomputer at NERSC. We note that any Parsl executor can be used in the program by using the appropriate configuration. For example, one can easily use a configuration for a specific machine, from their prior use of that machine, configurations published by HPC sites or from the Parsl documentation for many ACCESS and institutional HPC clusters.

\textbf{Creating CWLApps}:
As shown in listing \ref{lst:parslworkflow}, we create a Parsl CWLApp for each CommandLineTool by specifying the CWL file. 


\textbf{Defining the workflow}: We define the workflow by creating a Python function that includes the three stages. Each consumes the output of the prior stage, which establishes the dataflow graph. 

\textbf{Stage 1: Image Resizing}:
         The first stage uses the CWLApp, \texttt{resize\_image}, to resize each image to a specified size.
        A DataFuture is returned that is passed on to the next stage.

\textbf{Stage 2: Image Filtering}:
         The \texttt{filter\_image} CWLApp is used to apply a sepia tone filter to each resized image by setting the sepia input argument to True. The input image to this step is extracted from the DataFuture from the previous resizing step
         A new DataFuture is returned to reference the filtered image and is passed on to the next stage.

    \textbf{Stage 3: Image Blurring}:
         The final stage involves blurring each filtered image using the \texttt{blur\_image} CWLApp.
         The output images, obtained from the DataFutures of Stage 2, are used as input for for the blurring operation. 

    \textbf{Starting the workflow}: As the workflow is written in Python, we have access to the full capabilities of Python. In this case, we use a list comprehension and a glob pattern to identify all ``png'' files in the subdirectories and to start an instance of the workflow for each. We maintain a list of Futures for all processed images.

    \textbf{Wait for Results}:
         The use of Futures enables concurrent execution of the workflow stages for all images. 
         Parsl handles the workflow automatically, ensuring that the result of each future is available before executing the subsequent stage. 
         The pipeline waits for all processing futures to complete before concluding, ensuring all images are fully processed.

\subsection{Discussion}
We have shown how the image processing workflow can be implemented in Parsl by importing the CWL CommandLineTool definitions. As shown, the Parsl implementation offers an intuitive Python implementation using the full power of the Python programming language, enabling modular workflow definition, simple looping over inputs, asynchronous execution, and access to Parsl's executor and provider ecosystem. The same workflow can therefore be easily moved between computers and scaled from laptops to supercomputers. 

\section{Python Expressions in CWL Workflows}
\label{sec:inlinepython}

It is clear that static workflow representations are not suitably expressive for many scientific workflows. 
Indeed, such observations motivated the development of Parsl and it's predecessor Swift~\cite{swift}, while other static specification languages like CWL have evolved to support more dynamic behavior via inclusion of code.  






CWL supports specification of dynamic \textit{expressions} within workflow definitions. These expressions, written in JavaScript, enable CWL workflows to adapt their behavior dynamically. For example, expressions can be used to modify input arguments to pass to CommandLineTools, modify arguments passed between stages, or operate on results from CommandLineTools. Here, we propose a new Python-based expression that better matches the execution environment of Parsl. Further,  Python is a widely-used, easy-to-understand, and productive language. 

We approach this problem by defining a new type of expression: \textbf{InlinePythonRequirement}.
We follow closely the CWL \textit{InlineJavascriptRequirement} used for JavaScript.  We allow inclusion of Python code, including functions, to be specified in the InlinePythonRequirement section.  As a referenceable requirement, the expression can then be reused throughout the workflow definition. 

When invoking the Python expression, we need a way to refer to variable attributes in the workflow (e.g., input arguments) and to select the arguments to be passed to the expression. We adopt a simple notation for referencing attributes in the workflow: \texttt{\$(inputs.input)}.Here, the ``\$'' indicates the reference, ``inputs'' the input arguments, and ``input'' the specific argument.
We adopt a Python f-string-like~\cite{pep498} syntax to template arguments to be passed to the InlinePythonRequirement. This allows expressions to be used anywhere in the CWL workflow definition. 

\lstset{style=yamlstyle}
\begin{lstlisting}[caption={InlinePythonExpression capitalizing words}, label={lst:capital}]
cwlVersion: v1.2
class: CommandLineTool
requirements:
  - class: InlinePythonRequirement
    expressionLib:
      - |
        def capitalize_words(message):
            """
            Capitalize each word in the given message.

            Args:
                message (str): The input message.

            Returns:
                str: The message with each word capitalized.
            """
            return message.title()

baseCommand: echo

inputs:
  message:
    type: string

arguments: 
    - f"{capitalize_words($(inputs.message))}"

outputs: []
\end{lstlisting}


\autoref{lst:capital} and \autoref{lst:csv}  show examples of how the InlinePythonExpression can be used.  \autoref{lst:capital} shows a simple CommandLineTool that calls the echo command and uses the Python Expression to capitalize the result. Here the InlinePythonExpression defines a \textit{capitalize\_words} function to capitalize the input message (a string).  
The InlinePythonExpression is called in the argument attribute (arguments are used in CWL to create additional options or modify workflow inputs for invoking a CommandLineTool). The invocation of the \textit{capitalize\_words} function is enclosed within a Python `f-string', signaling to \textit{parsl-cwl} that it is a Python expression requiring evaluation. The argument processes the user input message, calls the InlinePythonExpression to capitalize the input, and then invokes the echo CommandLineTool with the capitalized message as input. 

InlinePythonExpressions enable the inclusion of Python expressions anywhere in the workflow, which is particularly useful for input validation.
Listing~\ref{lst:csv} shows this with the use of a 
\textit{validate} field for each input. In this example, the CWL CommandLineTool has a validate field that invokes a Python function \textit{valid\_file} to verify that the input data file is of type `.csv'. This validation is done before the execution of the CWL CommandLineTool allowing validation to be incorporated directly within the CWL file rather than this being handled externally.

\begin{lstlisting}[caption={InlinePythonExpression example to verify that a given input file is a CSV file}, label={lst:csv}]
cwlVersion: v1.2
class: CommandLineTool

requirements:
  - class: InlinePythonRequirement
    expressionLib: 
    - |
      def valid_file(file, ext):
          """
          Check if a file is valid

          Args:
              file (str): Path to the file
              ext (str): Expected file extension

          Raises:
              Exception: If the file is invalid
          """

          if not file.lower().endswith(ext):
              raise Exception(f"Invalid file. Expected '{ext}'")

baseCommand: cat

inputs:
  data_file:
    type: File
    validate: |
        f"{valid_file($(inputs.data_file), '.csv')}"
    inputBinding:
      position: 1

outputs:
  validated_output:
    type: stdout
\end{lstlisting}

\begin{figure*}[h!]
    \centering
    \begin{subfigure}[b]{0.45\textwidth} 
        \centering
        \includegraphics[scale=0.59]        {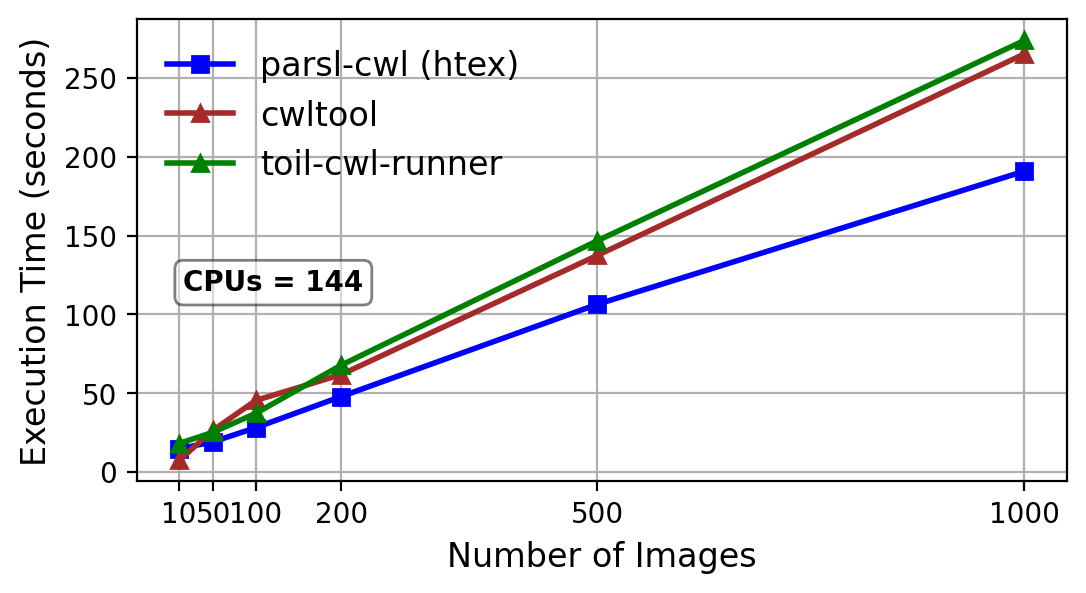}
        \caption{Three nodes}
        \label{fig:runtime_comparison1}
    \end{subfigure}
    \begin{subfigure}[b]{0.45\textwidth} 
        \centering
        \includegraphics[scale=0.59]{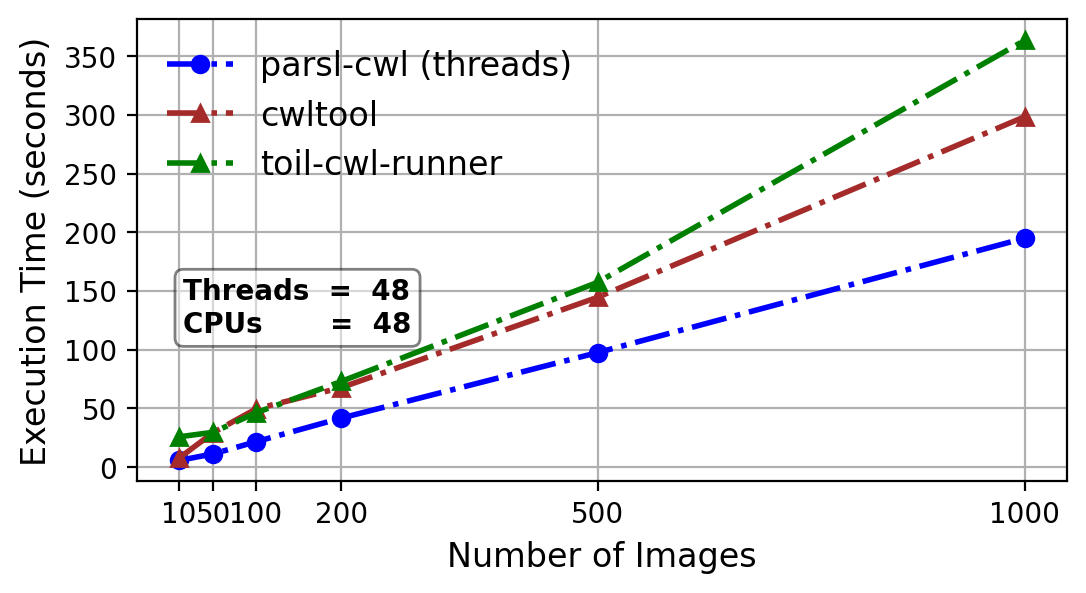}
        \caption{Single node}
        \label{fig:runtime_comparison2}
    \end{subfigure}
    \caption{Runtimes for CWL image processing workflow using CWLTool, Toil and Parsl-CWL on three nodes and one node}
    \label{fig:runtime_comparison_main}
\end{figure*}

These Python expressions can support multiple use cases:

\begin{itemize}
    \item \textbf{Dynamic Input Validation}: Python can be used to implement complex input validation logic, ensuring data integrity and preventings runtime errors. For example, fields can be checked for valid ranges, formats, or dependencies, with exceptions raised for invalid data.

    \item \textbf{Error Handling}: Python’s exception-handling mechanisms can be utilized to manage errors, enhancing workflow reliability.
        
    \item \textbf{Conditional Defaults}: Inline Python allows the specification of default values based on other inputs, enabling dynamic parameter management. This capability is particularly useful for workflows that require adaptive default values derived from related inputs.
    
    \item \textbf{External Python files}: External python files can be imported making the functions and variables available to be used in other parts of the CWL document.
\end{itemize}

\section{Evaluation}
\label{sec:evaluation}

The primary contribution of our work is the integration of CWL and Parsl and support for inline Python expressions in CWL.  
The examples presented illustrate the functional capabilities realized from this work. Here, we briefly explore the performance of our CWL and Parsl integration as well as Python expressions.

\subsubsection{CLW + Parsl}
We evaluate our integration by comparing execution of a CWL workflow using Parsl with ones executed using cwltool and Toil. 

We implement a CWL workflow using the image processing workflow detailed in Listing~\ref{lst:CWL_Image_Processing_Example} with a wrapper to process a list of images. The wrapped workflow uses the scatter method to call the sub workflow on each image individually. This approach ensures that the execution of the image processing workflow on each image is independent, allowing cwltool and Toil to leverage parallel execution of these independent steps.

We conducted our experiments on a high-performance computing cluster located in our department. We used two configurations: 1) a single-node configuration using local threads or processes; 2) a distributed configuration using three nodes. 
Each node in our cluster is equipped with two 12-core Intel x86\_64 processors (48 logical CPUs) and 126GB of RAM.
We configured cwltool with the \texttt{parallel} option and toil-cwl-runner with the \texttt{slurm} batch system. 
We configured the Parsl workflow using the example in Listing~\ref{lst:parslworkflow} and using the HighThroughputExecutor, for three nodes, and ThreadPoolExecutor for the single-node deployment. 
We configured each workflow system to use all cores available on the allocated nodes.



Figure~\ref{fig:runtime_comparison_main} shows a linear scaling trend as the number of images increases in both three node and single node deployments. Using three nodes (Figure~\ref{fig:runtime_comparison1}), Parsl-CWL with the HighThroughputExecutor achieves approximately 1.5 times better performance than CWLTool when processing a workload of 1,000 images. Similarly, on a single node (Figure~\ref{fig:runtime_comparison2}), Parsl-CWL using the ThreadPoolExecutor outperforms CWLTool by about the same factor for the same workload.



\subsubsection{Python Expressions in CWL}

We now consider the time to evaluate InlinePythonExpressions compared to CWL's InlineJavaScriptExpressions. We use the simple workflow from Listing~\ref{lst:capital} where the expression simply changes the case of a set of words. We deploy the workflow on a single node on the HPC cluster.  We scale the number of words and record the time to complete the workflow. In
\figurename~\ref{fig:runtime_comparison3}, we see a short time to process up to 1024 words using the InlinePythonExpression, a constant performance for the Inline Python Expressions. We see a superlinear increase in time for JavaScript Expressions using both cwltool and Toil. 

\begin{figure}[h]
    \centering
    \includegraphics[scale=0.60]{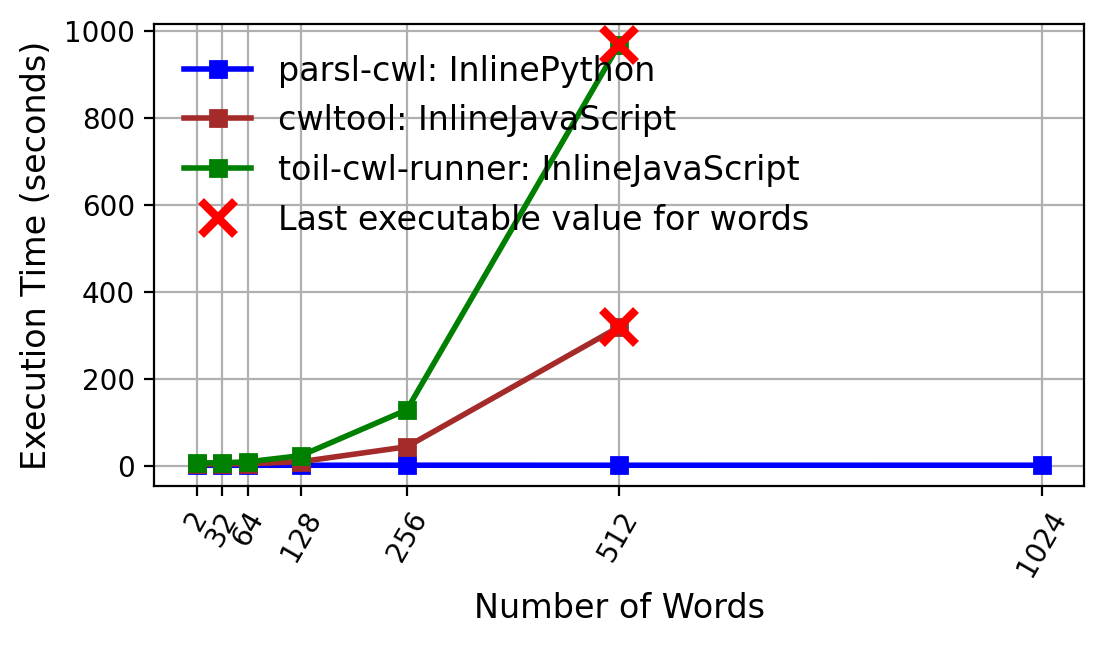}
    \caption{Runtime for CWL InlineJavaScript processing using CWLTool, Toil and InlinePython using Parsl-CWL as we increase number of words from 2 to 1024}
    \label{fig:runtime_comparison3}
\end{figure}

\section{Related Work}
\label{sec:relatedwork}


Many workflow systems---353 at the 
time of writing~\cite{workflowsystems}---enable the orchestrated execution of multiple applications at both small and large-scale. Well-known examples include Pegasus~\cite{pegasus}, Galaxy~\cite{galaxy}, Swift~\cite{swift}, RCT~\cite{balasubramanian2019radicalcybertools},  NextFlow~\cite{di2017nextflow}, FireWorks~\cite{jain2015fireworks}, Apache Airflow~\cite{airflow}, and Luigi~\cite{luigi}.
These systems are differentiated by their language, 
static workflow definition, explicit graph definition, 
or focus on specific domains or communities. 
Community workflow specifications such as CWL and WDL~\cite{wdl} have become increasingly popular; however, neither WDL or CWL are themselves workflow engines, and instead they may be executed using one of several supported workflow engines. 

CWL has developed an extensive ecosystem of runners. The reference implementation CWLTool and Toil~\cite{Vivian2017} are both implemented in Python. 
However, while they may be written in Python, the runner itself does not expose the Python side of the tool, instead providing a command line interface for users to run CWL workflows. 

To the best of our knowledge, JavaScript is the only language supported by CWL expressions. Others have developed tools in Python and Java to support creation of CWL definitions directly from other languages and to provide parsing and validation of CWL definitions~\cite{cwljava}. 

This paper builds on previous ideas, such as the concept of using common definitions for applications. For example, a previous paper~\cite{Stubbs2020} that included some of the authors of this paper suggested common configurations for both applications and systems, with the idea that the owners of those applications (developers) and systems (system administrators) could do a small amount of work to make these resources easy for researchers to use in their workflows.




\section{Summary}
\label{sec:summary}

The integration of CWL and Parsl enables Parsl workflows, written in Python, to easily leverage the ecosystem of CWL CommandLineTool definitions.  It allows users to combine these two environments, allowing CWL CommandLineTools to be run in the flexible and scalable Parsl environment on a variety of computing resources. 
Our performance evaluation shows that Parsl can execute CWL CommandLineTools efficiently and comparably to existing CWL runners. 
To bring CWL towards the Python ecosystem we presented a prototype integration of Python Expressions within CWL workflow definitions. Our approach enhances workflow flexibility and expressiveness, enabling researchers to implement complex logic, validate inputs, and manage dependencies directly within their workflows using Python. 
Our future work focuses on developing our prototype into a robust toolkit for Parsl and adding support in Parsl to run complete CWL workflows. 

\section{Acknowledgements}
This work was supported in part by the National Science Foundation awards 2209919 and 2209920, the Chan Zuckerberg Initiative, and the U.S. Department of Energy
under Contract DE-AC02-06CH11357.

\bibliographystyle{IEEEtran}
\bibliography{refs}

\end{document}